\documentclass{article}
\usepackage[utf8]{inputenc}
\usepackage{amsmath} 
\usepackage{amsthm}
\usepackage{mathtools}
\usepackage{enumerate}
\usepackage{amssymb}
\usepackage{bbm}
\usepackage{xcolor}
\usepackage[round]{natbib}
\usepackage{graphicx}
\usepackage{hyperref}
\usepackage{empheq}
\usepackage{geometry}
\usepackage{caption}
\usepackage{subcaption}
\usepackage{booktabs}
\usepackage{makecell}

\theoremstyle{plain}

\theoremstyle{definition}

\theoremstyle{remark}

\numberwithin{equation}{section}
\numberwithin{theorem}{section}

\newcommand{\E}{\mathbb{E}}

\begin{document}

\title{Optimizing Broker Performance Evaluation through\\Intraday Modeling of Execution Cost\thanks{Fruitful discussions with Umut Cetin and the research assistance of Zhengnan ``Christine'' Dong in the early stages of the project are gratefully acknowledged. We are also grateful for the generous feedback of Fengpei Li, Roel Oomen, Kevin Webster and Nicholas Westray.}}
\author{Zolt\'an Eisler\thanks{Imperial College London, Department of Mathematics} \thanks{Corresponding Author. Email: z.eisler@imperial.ac.uk.}
\and Johannes Muhle-Karbe\footnotemark[2]}

\maketitle
\begin{abstract}
Minimizing execution costs for large orders is a fundamental challenge in finance. Firms often depend on brokers to manage their trades due to limited internal resources for optimizing trading strategies. This paper presents a methodology for evaluating the effectiveness of broker execution algorithms using trading data. We focus on two primary cost components: a linear cost that quantifies short-term execution quality and a quadratic cost associated with the price impact of trades. Using a model with transient price impact, we derive analytical formulas for estimating these costs. Furthermore, we enhance estimation accuracy by introducing novel methods such as weighting price changes based on their expected impact content. Our results demonstrate substantial improvements in estimating both linear and impact costs, providing a robust and efficient framework for selecting the most cost-effective brokers.
\end{abstract}


\section{Introduction}

Minimizing trading costs in the execution of large orders is a central problem in financial theory and practice. Starting from the seminal papers of \citet{bertsimas1998optimal,almgren2001optimal,obizhaeva2013optimal}, this has given rise to large and active literature that derives the optimal execution paths for a variety of different price impact models (cf., e.g., the textbooks \citet*{cartea2015algorithmic,gueant2016financial,webster2023handbook} and the references therein).

However, many firms do not optimize their trading decisions themselves, opting instead to outsource their execution algorithms to specialized brokers. These brokers then provide their clients with price and trade data \textit{ex post}. The clients' key challenge, instead of fine-tuning how orders are sent, becomes picking the right broker. Automated broker selection algorithms are often called ``algo wheels". They allocate order flow to different brokers with the dual aim of minimizing trading costs right away, and collecting further data about each broker's performance to facilitate the selection later on. This is an exploration-exploitation problem, extensively studied in the literature on multi-armed bandits. 
A growing number of providers propose algo wheels based those principles, and take-up by clients is on the rise \citep{virtu2019algo, greenwich2019trends, bacidore2020algorithmic}.

This selection process crucially depends on the precise estimation of each broker's execution performance. The standard approach to this is to regress realized slippage (trading cost per unit quantity) against order size (or some power thereof) plus an intercept, leading to estimates for price impact and linear costs, respectively. This ``static’’ approach is simple and essentially model free, but the corresponding signal to noise ratio is typically small. Therefore large datasets of thousands of orders are required to obtain a reliable statistical model. 

To address this, our study outlines how to exploit the transient nature of price impact to improve the signal-to-noise ratios considerably.\footnote{A complementary approach to improve the quality of cost estimations is the ``causal regularization'' proposed by~\cite{webster2023getting}, which reduces the biases that arise when impact and alpha confound each other.} To estimate linear costs, this can be achieved by replacing slippage relative to the ``arrival price’’ with slippage relative to mid prices continuously sampled during the execution. We show that by doing so, one filters out a large part of impact cost and market noise, while preserving the linear cost term.\footnote{This is similar in spirit to the use of control variates in Monte Carlo simulation.} For impact costs, suitably re-weighting mid price changes with a greater emphasis on earlier rather than later trades leads to similar improvements.\footnote{This corroborates the recent findings of~\cite*{li2022price}, who show that the asymptotic efficiency of price impact estimates can be improved by utilizing the information of earlier trades.} The intuition here is that price impact has a much stronger effect on prices at the beginning of execution, but then gradually saturates~\citep*{durin2023two}.

We illustrate the underlying mechanisms and quantify the magnitude of the improvements that can be achieved in this manner for the standard price impact model of \citet{obizhaeva2013optimal}, and broker trades with simple linear dynamics oscillating around a TWAP baseline. These fluctuations are a reduced form representation of various factors like micro-scale alphas not communicated to client even ex post. For model parameters typical for the trading of E-mini S\&P futures, we find that our simple alternative estimators increase the signal-to-noise ratio by a factor of $6$ to $7$. Such an improvement can significantly speed up the convergence of broker selection procedures that are ubiquitous in the industry. While these specific improvements are derived in a particular model, the mechanisms underlying them are simple and robust. We therefore expect the broad conclusions to remain valid for a large range of other models for the trading process and impact.

The remainder of this article is organized as follows. Section \ref{sec:setting} introduces our general framework for modeling TWAP-like trading and the resulting price dynamics. Section~\ref{sec:naive} in turn establishes a baseline by calculating the mean and variance of the execution costs. Section~\ref{sec:linear} then proposes an improved estimator for the linear cost in our model, and shows that the variance is greatly reduced compared to the na\"ive approach. Section~\ref{sec:impact} defines another estimator that achieves a similar increase in signal-to-noise ratio for the impact cost. Finally, \ref{sec:discussion} presents some concluding remarks and outlook. The detailed derivations of the formulas reported in this paper are collected in an online appendix~\citep{appendix}.

\section{Setting the Stage}

\label{sec:setting}

We consider a client who consistently places orders for a particular asset through a broker.\footnote{Our approach can also be applied in various analogous settings, such as for the portfolio of a hedge fund and its in-house execution algorithms.} In this section, we introduce a simple model for these trades, their impact on market prices, and the method by which clients seek to quantify the relationship between the two.

\subsection{Broker Trading}

We assume that the broker executes the orders over some fixed time period $t\in[0, T]$ with a  trading rate $q_t$, representing the broker's adjustments of the position $Q_t = \int_0^t q_s ds$. This process can be observed ex post by the client, based on trading reports.

Typical practical examples are TWAP algorithms (for which the trading rate is constant) and VWAP algorithms (where the TWAP rate is adjusted with the realized trading volume). To simplify the exposition, we focus on the simpler TWAP benchmark in this paper, but allow the trading speed to fluctuate with Ornstein-Uhlenbeck (OU) dynamics: 
\begin{equation}
dq_t = \frac{1}{\tau_q} \left(\frac{Q}{T}-q_t\right)dt+ \frac{Q}{T} \sigma_q dW^q_t. 
\label{eq:OU}
\end{equation}
Here, $Q$ is the total order size and $T>0$ is the time horizon over which it would be executed with the TWAP rate $Q/T$. The parameters $\sigma_q>0$ and $\tau_q>0$ describe the magnitude of the fluctuations around the TWAP baseline, and the timescale over which these decay.

This model is simpler, but similar in spirit to the one in~\cite[Chapter~9.2]{cartea2015algorithmic}. To wit, the random fluctuations describe, in reduced form, that the broker may deviate from the target schedule to take advantage of short-term alpha signals or local changes of liquidity, for example. To further simplify the analysis, we focus on the stationary version of the OU process~\eqref{eq:OU}, for which $q_t$ follows a normal distribution with constant mean $\mathbb{E}[q_t] = Q/T$ and variance $\mathrm{Var}[q_t] = Q^2\sigma_q^2 \tau_q/2T^2$. The expected total quantity executed by the broker in turn becomes $\mathbb{E}[Q_T] = Q$, in line with the TWAP benchmark. The latter is recovered exactly in the limit $\tau_q \to 0$, for which the fluctuations around the TWAP rate $Q/T$ disappear.

\begin{figure}[tb]
\centering
\begin{subfigure}{.45\textwidth}
  \centering
  \includegraphics[width=0.99\textwidth]{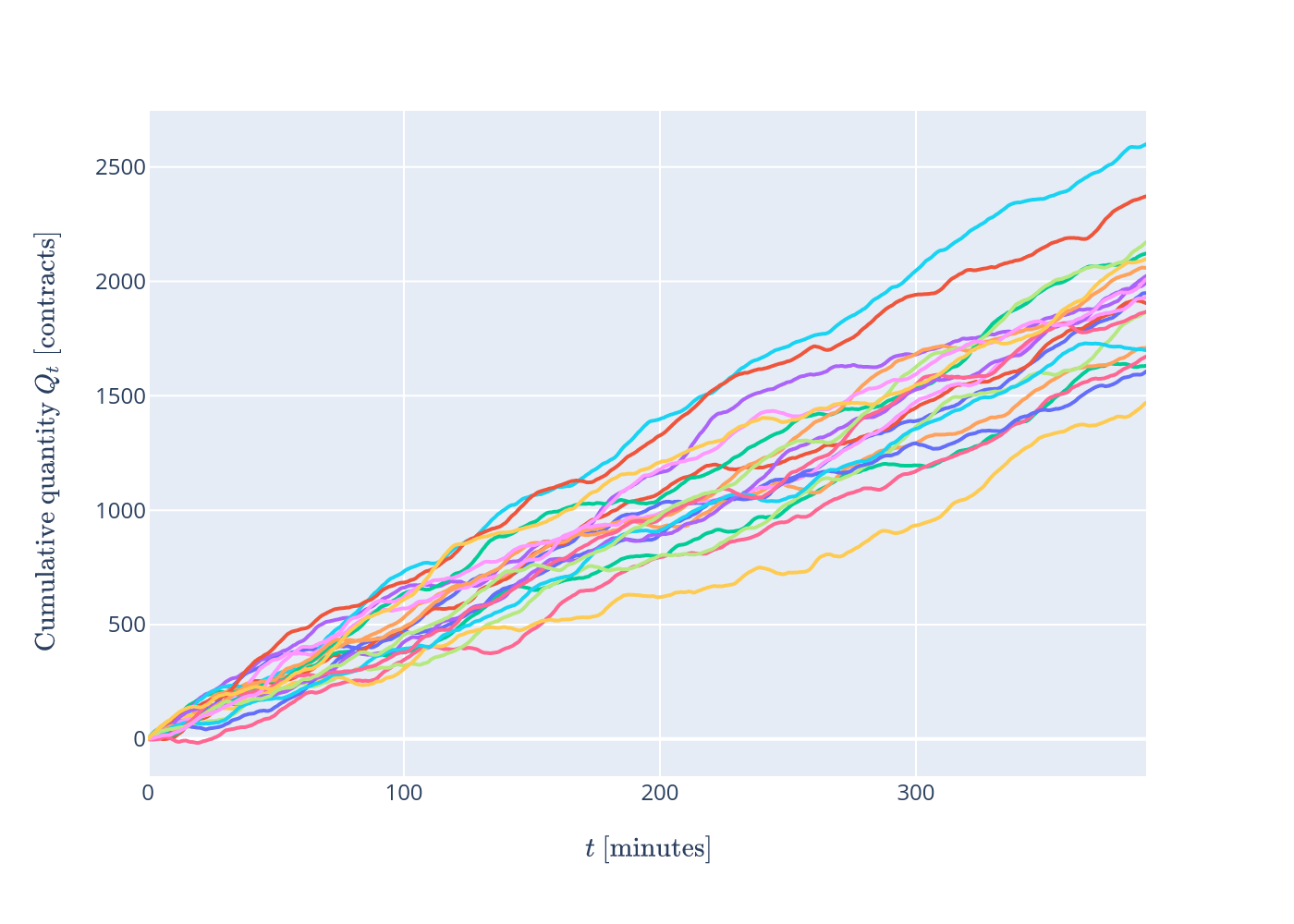}
  \caption{Sample paths for the unconditional process.}
  \label{fig:sub1}
\end{subfigure}
\begin{subfigure}{.45\textwidth}
  \centering
  \includegraphics[width=0.99\textwidth]{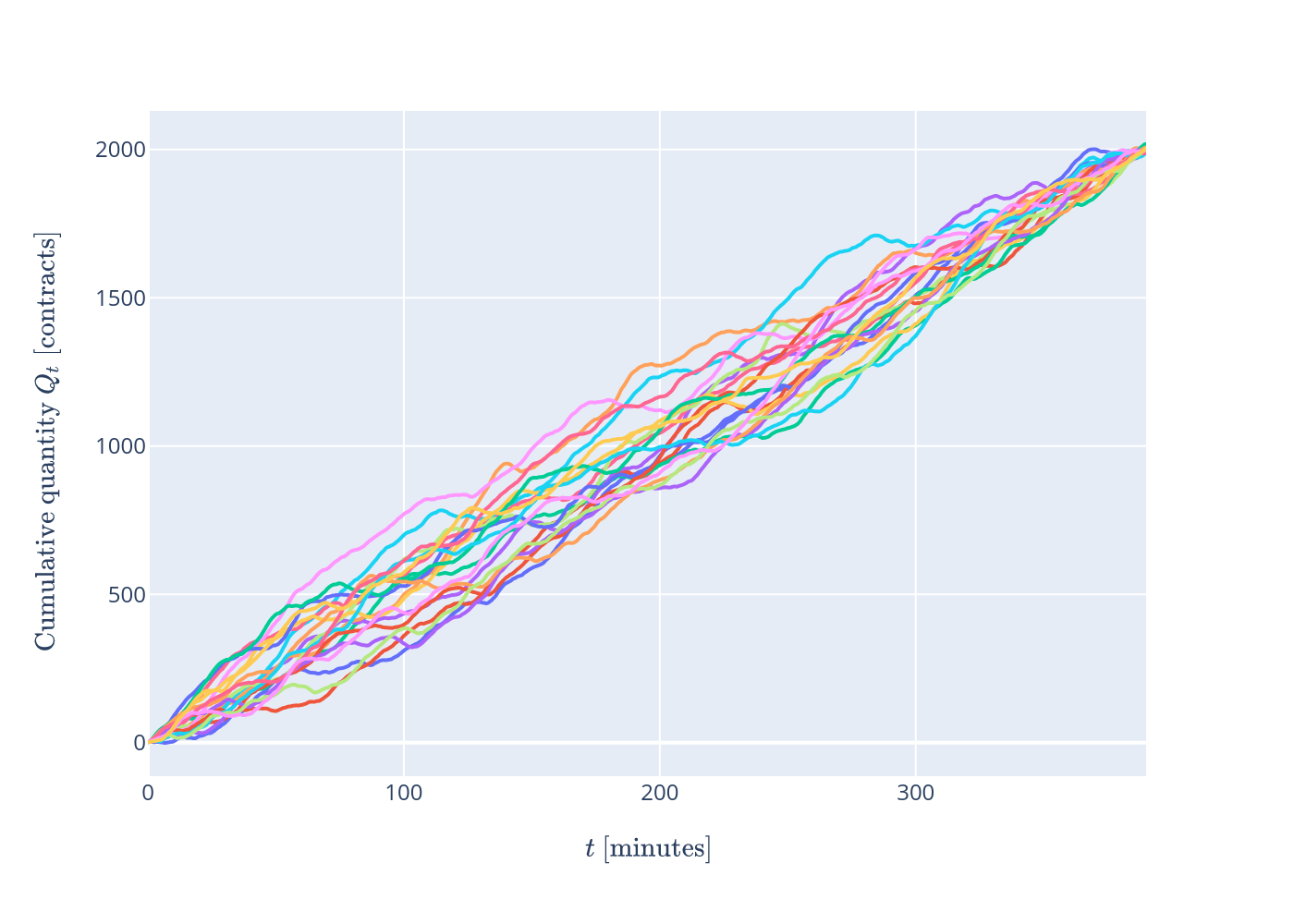}
  \caption{Sample paths for the conditional process.}
  \label{fig:sub2}
\end{subfigure}
\caption{Twenty randomly selected sample paths of the cumulative quantity $Q_t$ to illustrate the typical shape and variability of trading profiles.}
\label{fig:cumulative_quantity}
\end{figure}

Perfect execution of the target quantity $Q$ can be enforced by conditioning on $Q_T=Q$. Some simulated sample paths of $Q_t$ based on a realistic parameter set (cf.~Table~\ref{tab:table_params}) are shown for both the conditional and the unconditional process in Fig.~\ref{fig:cumulative_quantity}. In view of the SDE representations derived by~\citet{cetin.danilova.16}, this conditioning leads to a modified linear process whose dynamics are still Gaussian. As a consequence, most of our analysis could be extended to this setting. In order to avoid drowning our main points in complicated algebra, we nevertheless focus on the simpler unconditional model~\eqref{eq:OU}. Later on, we verify by numerical simulation that similar results are obtained for its conditional counterpart.

\subsection{Price Impact}

Price impact is the causal relation between the broker's trades and the \emph{mid price} $M_t$. We describe this by the standard price impact model of~\citet{obizhaeva2013optimal}:\footnote{Note that the TWAP baseline we assume mirrors the (risk-neutral) optimal execution path in the model of ~\citet{obizhaeva2013optimal} on the interior $(0,T)$ of the trading interval. We disregard the initial and terminal bulk trades that are optimal in this model to keep the model as simple as possible, but still sufficiently faithful to the real-life behavior of brokers.}
  \begin{eqnarray}
        M_t - M_0 = \lambda \int_0^t e^{-(t-t')/\tau_M}q_{t'}dt' + \sigma_M W^M_{t}.
        \label{eq:mtm0}
    \end{eqnarray}
This means that the unaffected price follows a Brownian motion $W^M_t$ with volatility $\sigma_M>0$. We assume it is independent from the Brownian motion $W^q_t$ driving the fluctuations of the broker's trading rate. The broker's trades in turn have linear price impact proportional to $\lambda>0$, often referred to as ``Kyle's lambda''~\citep{kyle1985continuous}. Impact subsequently decays at rate $1/\tau_M>0$, i.e., the half-life of impact is proportional to $\tau_M$.

The \emph{execution price} $P_t$ achieved by the broker additionally includes a fraction $a$ of the bid-ask spread $s>0$:
\begin{equation}
    P_t = M_t + as\times\mathrm{sign}(q_t).
    \label{eq:pt}    
\end{equation}
To simplify the analysis, we assume that both the impact decay timescale $\tau_M$ and the spread $s$ are known constants.  The quality of a broker is in turn determined by the remaining two effective parameters:
\begin{itemize}
    \item The constant $\lambda$ quantifies the amount of price impact caused by the broker's trades. A high value will make larger orders excessively expensive, as repeated trading creates consecutive adverse price moves.
    \item The parameter $a$ captures the short-term execution quality such as capture of the bid-ask spread, short-term alpha and adverse selection, among other factors. As a worst-case scenario, if a broker always aggressively crosses the spread to take liquidity, and has no skill in timing the orders to before favorable price moves, one expects $a=1/2$. In practice the value of $a$ is lower in most other cases.
\end{itemize}

\subsection{The Client's Inference Problem}

The optimal execution problem involves finding the most cost-effective trading schedule $q_t$, given the order size $Q$. The values of the cost parameters $a$ and $\lambda$ are either given, or also have to be estimated from data. This is a well-researched area in finance (cf., e.g., \cite{cartea2015algorithmic,gueant2016financial,webster2023handbook} and the references therein). In the present study, we focus on a different scenario known as broker evaluation~\cite{webster2023handbook}. Here, the client lacks knowledge of $\lambda$ and $a$, and aims to estimate these parameters from trading data. While the client decides on the order size $Q$, the specific trading schedule $q_t$ is not chosen, but only observed. This last point is the crucial difference of our approach compared with an optimal execution setting. For more context, we refer the reader to the works of \citet{almgren2005direct, bershova2013non, toth2011anomalous}.

Once the total quantity $Q$ is selected, the broker has discretion over the trading process. To estimate the parameters $a$ and $\lambda$, institutional grade executing brokers readily offer data such as the realized trading schedule $q_t$, the resulting fill prices $P_t$, and the time evolution of the mid-price $M_t$. Eqs.~\eqref{eq:mtm0} and~\eqref{eq:pt} link the observed prices to these unknown parameters and the trading schedule.

The challenge in inference arises because while $q_t$ is directly observed, $P_t$ and $M_t$ are strongly affected by market noise, complicating the estimation of $\lambda$ and $a$. Our goal is to minimize the resulting statistical error by leveraging the available data efficiently. This is important, because more accurate parameter estimates for different brokers enable the client to better predict the most cost-effective broker for a given trade size. 

\subsection{Parameter Values}

In our numerical examples, we select sensible parameter values for trading E-mini S\&P futures, see Table~\ref{tab:table_params}. To wit, the target trade size $Q$ is taken to be roughly $0.2\%$ of the real-life trading volume which is of the order of $10^6$ contracts. The typical price impact of trading such an amount is a few percent of daily volatility. For the impact decay $\tau_M$ we choose a fairly short time scale to reflect that the turnover of E-minis is very high. 

The parameters for broker behavior ($\sigma_q$ and $\tau_q$) cannot be precisely inferred from public data, and may show some variation across market participants. However, the typical order of magnitude is to allow $Q_t$ to deviate from the TWAP value $tQ/T$ by a few percent of $Q$ at any time. This is supported by our own experience, as well further anecdotal evidence in \citet{bacidore2020algorithmic}. \citet[Section 9.2]{cartea2015algorithmic} calibrate an optimal execution model to market data, and reach similar conclusions. 

\begin{table}[h]
\centering
\caption{Model parameters for the numerical simulations. They represent sensible values for trading E-mini S\&P futures. The bid-ask spread corresponds to $2$ bps of the price. Volatility is normalized such that over the course of the order the mid price typically varies by $1.0\%$.}
\label{tab:table_params}
\begin{tabular}{rrrrr}
\toprule
 & Parameter value \\
\midrule
$M_0$ & $5000.0$ \\
$Q$ & $2000$ contracts \\
$T$ & $390$ minutes \\
$\tau_M$ & $39$ minutes \\
$\tau_q$ & $5$ minutes \\
$a$ & $0.5$ \\
$\lambda$ & $0.0075$ \\
$\sigma_M$ & ${2.532}/$minute$^{0.5}$ \\
$\sigma_q$ & $0.5$ \\
$s$ & $1.0$ \\
\bottomrule
\end{tabular}
\end{table}

\section{Na\"ive Estimation of Execution Cost}

\label{sec:naive}

As a reference point, we first consider the standard approach of estimating trading costs from the ``slippage'' relative to the ``arrival price'' $M_0$:
\begin{eqnarray*}
C_T = \int_0^T (P_t - M_0) q_t dt.
\end{eqnarray*}
In the Obizhaeva-Wang model, this can be decomposed as 
\begin{equation}\label{eq:Cost}
C_T=\underbrace{as \int_0^T |q_t| dt}_\mathrm{linear\ (spread)\ cost} + \underbrace{\lambda \int_0^T \left(\int_0^t e^{-(t-s)/\tau_M} q_s ds\right)q_t  dt}_\mathrm{impact\ cost} + \underbrace{\sigma_M \int_0^TW^M_t q_t dt}_\mathrm{market\ noise}.
\end{equation}
Throughout this paper we will assume that the half-lives of impact and trading speed fluctuations are short relative to the length $T$ of the trading interval: 
$$\tau_M,\tau_q=O(\tau) \quad \mbox{and in turn also} \quad  \mathrm{Var}[q_t]=O(\tau).
$$

\subsection{Linear Cost}

For the expected linear cost in~\eqref{eq:Cost}, fluctuations of the trading rate around the TWAP baseline average out at the leading order:
\begin{align*}
\mathbb{E}[C_T^\mathrm{linear}] = as \int_0^T \mathbb{E}[|q_t|] dt  &= asQ + O(e^{-1/\tau}).
\end{align*}
For realistic parameters -- such as the ones from Table~\ref{tab:table_params} -- the correction term is negligible. As a result, linear costs are typically indistinguishable from the ones for TWAP trading, see Table~\ref{tab:table_results}.

\subsection{Impact Cost}

The expected impact cost in Eq. ~\eqref{eq:Cost} can be computed as~\citep{appendix} 
\begin{equation}\label{eq:ECTimpact}
\begin{split}
\mathbb{E}[C_T^\mathrm{impact}]
=&\underbrace{\lambda Q^2\frac{\tau_M}{T}\left(1-\frac{\tau_M}{T}\right)}_\mathrm{average\ TWAP\ impact\ cost}+\underbrace{\lambda \mathrm{Var}[Q_T]\frac{\tau_M}{T}\left(1-\frac{\tau_M}{T}\right)}_{\mathrm{correction\ for\ variability\ of\ } Q_T}\\&+\underbrace{\lambda Q^2\frac{\mathrm{Var}[q_t]}{E[q_t]^2}\frac{\tau_M \tau_q(T-2(\tau_M+\tau_q))}{(\tau_M+\tau_q)T^2}}_\mathrm{concentration\ penalty\ and\ residuals} + O(\tau^3).
\end{split}
\end{equation}

Again, the first term (of order $O(\tau)$) is the impact cost for a TWAP order. The following two terms (of order $O(\tau^2)$) represent the leading-order corrections to the impact cost due to fluctuations around the TWAP benchmark. These corrections originate from two related effects. First, in the unconditional model trading speed fluctuations add up to a total executed quantity $Q_T$ that is generally different from the target quantity $Q$ for each order. The impact cost of a perfect TWAP order is quadratic in the quantity. Due to this convex relationship, the variability of $Q_T$ in turn induces this positive correction. Second, the fluctuations around the baseline trading rate momentarily over-/under-concentrate trading, which also alters impact costs. This fluctuation of the intraday profile adds another positive correction in the realistic regime $T>2(\tau_q+\tau_M)$.

\subsection{Estimation of Cost Parameters}

\label{ss:naive}

The market noise term has expectation zero, because the Brownian motions driving unaffected prices and broker trades are independent. The aggregate expected cost in turn is
\begin{equation}\label{eq:ECT}
    \begin{split}
    \mathbb{E}[C_T] &= asQ + \lambda Q^2\frac{\tau_M}{T}\left(1-\frac{\tau_M}{T}\right)+\lambda\mathrm{Var}[Q_T]\frac{\tau_M}{T}\left(1-\frac{\tau_M}{T}\right) \\ &\quad  +\lambda Q^2\frac{\mathrm{Var}[q_t]}{E[q_t]^2}\frac{\tau_M \tau_q(T-2(\tau_M+\tau_q))}{(\tau_M+\tau_q)T^2} + O(\tau^3).
    \end{split}
\end{equation}
How do clients typically estimate the cost parameters $a$ and $\lambda$? After having sent orders of varying sizes $Q$, they analyze the dependence of the measured costs (per unit traded) on that trade size. The client's inference problem in turn can be recast as the linear regression
\begin{align}
    C_T/Q = a\phi_1 + \lambda\phi_2 Q+ \epsilon.
    \label{eq:regression}
\end{align}
After substituting the values of $\phi_1$ and $\phi_2$ from Eq.~\eqref{eq:ECT}, the trading cost parameters $a$ and $\lambda$ can be determined from the intercept and slope of a regression of the slippage $C_T/Q$ against the trade size $Q$. 

The corresponding noise level $\epsilon$ is in turn determined by the variance of the execution costs. For TWAP trading, both the linear and the impact cost are deterministic and therefore do not contribute to the variance. The latter is simply given by the variance induced by changes of the unaffected price:\footnote{Here, we have used in the last step that the time integral $\int_0^T W_t dt$ of Brownian motion is Gaussian with mean zero and variance $T^3/3$.}
\begin{align*}
\mathrm{Var}[C^{\mathrm{TWAP}}_T]= \mathrm{Var}\left[ \int_0^T \sigma_M W_t \frac{Q}{T} dt \right]= \frac{\sigma_M^2 Q^2 T}{3}.
\end{align*}

In the general case, when the trading rate is also random, both the linear and quadratic trading costs also become random, and therefore further increase the overall variance. However, if price impact and fluctuations around the TWAP rate both decay quickly, then this only leads to small correction terms~\citep{appendix}:
\begin{equation}\label{eq:VarCT}
\begin{split}
\mathrm{Var}[C_T] = &\frac{\sigma_M^2Q^2T}{3}+\frac{\sigma_M^2\mathrm{Var}[Q_T]T}{2} +a^2 s^2\mathrm{Var}[Q_T] \\
&\quad -asQ\times \frac{\lambda\mathrm{Var}[Q_T]\tau_M}{T}\times\frac{T}{\tau_M+\tau_q}+O(\tau^3). 
\end{split}
\end{equation}
In Table \ref{tab:table_results} we compare these analytical formulas for the mean and variance with numerical simulations. We find that our approximation is quite accurate, and that the means are much lower than the corresponding standard deviations. This means that when estimating linear and impact costs via the regression~\eqref{eq:regression}, we need a very large sample of orders to obtain accurate results. For example, for a realistic sample of $N=1,000$ orders, the t-statistic of the linear cost term is only $\sqrt{N}\times \mathbb{E}[C_T^\mathrm{linear}]/\sqrt{\mathrm{Var}[C_T]} \approx 0.54$. For the impact cost the t-statistic is $\sqrt{N}\times \mathbb{E}[C_T^\mathrm{impact}]/\sqrt{\mathrm{Var}[C_T]}\approx 1.58$. Numerical simulations reported in Table~\ref{tab:table_results} confirm very similar results obtained for the conditional version of the model (where exact execution of the target quantity is enforced).

\begin{table}[h]
\centering
\caption{Numerical comparison of the analytical and simulation results for various quantities. Costs are expressed in US dollars, and impacts in price points. In the calculation of costs we took into account that 1 point of price variation corresponds to \$50 of gain/loss on a single contract.}
\label{tab:table_results}
\begin{tabular}{rrrrr}
\toprule
 & \makecell{Analytical\\leading order} & \makecell{Analytical\\full expression} & \makecell{Unconditional\\simulation} & \makecell{Conditional\\simulation} \\
\midrule
$\mathbb{E}[C^\mathrm{linear}_T]$ &  & \$50,000 & \$50,002 & \$50,013 \\
$\mathbb{E}[C^\mathrm{impact}_T]$ & \$135,000 & \$145,413 & \$145,641 & \$143,876 \\
$\mathbb{E}[\Delta C_T]$ & \$50,000 & \$60,653 & \$60,820 & \$58,331 \\
$\sqrt{\mathrm{Var}[C_T]}$ & \$2,886,751 & \$2,921,066 & \$2,919,396 & \$2,896,045 \\
$\sqrt{\mathrm{Var}[\Delta C_T]}$ & \$447,572 & \$447,617 & \$443,427 & \$248,747 \\
$\mathbb{E}[I]$ &  & 1.5 & 1.5 & 1.433 \\
$\sqrt{\mathrm{Var}[I]}$ & 50.0 & 50.029 & 49.976 & 49.97 \\
$\mathbb{E}[I_\pi]$ & 12.324 & 12.299 & 11.085 & 11.138 \\
$\sqrt{\mathrm{Var}[I_\pi]}$ &  & 50.0 & 47.435 & 47.681 \\
sample size &  &  & 10,000,000 & 1,000,000 \\
\bottomrule
\end{tabular}
\end{table}

\section{Enhanced Estimation of Linear Cost}

\label{sec:linear}

We now show how to substantially increase the signal-to-nose ratio by using more efficient statistics than the slippage relative to the arrival price. We first focus on the estimation of the linear cost parameter $a$. To this end, we compare the cost of each trading trajectory to the time-weighted average of the corresponding mid prices:
\begin{align*}
\Delta C_T &=\int_0^T (P_t-P_0)q_t dt-\int_0^T (M_t-M_0)\frac{Q}{T}dt\\
&=as \int_0^T q_t dt +\int_0^T \left(\left(\int_0^t \lambda e^{-(t-s)/\tau_M}q_s ds\right)\left(q_t-\frac{Q}{T}\right)\right)dt+\int_0^T \sigma_M W^M_t \left(q_t-\frac{Q}{T}\right)dt.
\end{align*}
The key here is that in calculating the time weighted average in the reference strategy, we use the actual realized mid prices, which were themselves affected by both the price impact of real trading and the market noise. We therefore eliminate most of those two components and are left, to an extent much greater than before,  with the linear cost. 

Reusing results for the expected cost from the previous section, we find
\begin{align}
\E[\Delta C_T/Q]=as + \lambda Q \frac{\mathrm{Var}[q_t]}{E[q_t]^2}\frac{\tau_M \tau_q}{(\tau_M +\tau_q)T}+O(\tau^3).
\label{eq:EDeltaCT}
\end{align}
The corresponding variance can also be obtained~\citep{appendix}:
\begin{align}
\mathrm{Var}[\Delta C_T/Q] &= \underbrace{\sigma_M^2\frac{\mathrm{Var}[q_t]}{\mathbb{E}[q_t]^2} \tau_q}_{\mathrm{market\ noise}} + \underbrace{a^2s^2\frac{\mathrm{Var}[Q_T]}{Q^2}}_{\mathrm{correction\ for\ variability\ of\ } Q_T} + O(\tau^3).
\label{eq:VarDeltaCT}
\end{align}
In what sense is regressing $\Delta C_T/Q$ on $Q$ with an intercept a better estimator than it's na\"ive counterpart~\eqref{eq:regression} from Section~\ref{ss:naive}? Comparing the expectation in Eq.~\eqref{eq:EDeltaCT} with that of the na\"ive cost in Eq.~\eqref{eq:ECT} we see that, as we hoped, the linear cost term is still present and unchanged. However, most of the quadratic impact cost terms are eliminated, except for the highest-order correction.

Crucially, the corresponding variance is also much smaller. Indeed, comparing Eqs.~\eqref{eq:VarCT} and \eqref{eq:VarDeltaCT}, we see that only two terms remain. The market noise changes from $\sigma_M^2\mathbb{E}[Q_T]^2T/3$ to $\sigma_M^2\mathrm{Var}[q_t] \tau_q T^2$, a 6.5 times reduction with our example parameters. The second -- small -- contribution is due to the variability of realized trade size $Q_T$ in the unconditional version of the model. For 1,000 orders the t-statistic of linear cost in turn increases from the value $0.54$ of the na\"ive approach to $\sqrt{N}\times \mathbb{E}[C_T^\mathrm{linear}]/\sqrt{\mathrm{Var}[\Delta C_T]} \approx 3.57$ in the unconditional version of the model, and to even $7.42$ in the conditional version of the model with $Q_T=Q$.

\begin{figure}[tb]
    \centering
    \includegraphics[width=0.8\linewidth]{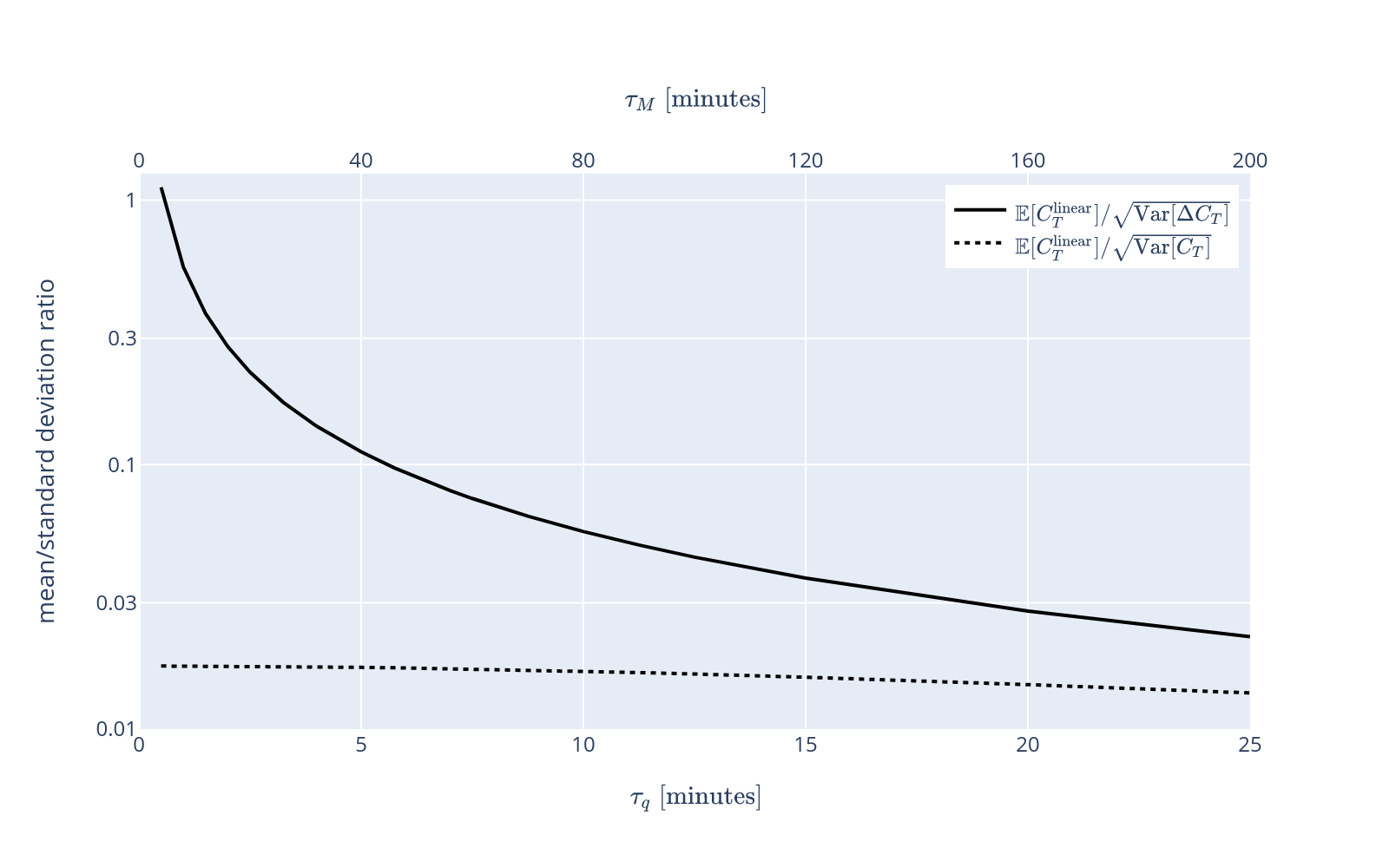}
    \caption{The ratio of the mean and standard deviation of the two linear cost metrics calculated from regression intercepts. The dotted line represents the na\"ive metric $C_T$, cost to arrival price. The solid line represents $\Delta C_T$, cost to TWAP. On the horizontal axis we vary in tandem the time scales of trading rate fluctuations ($\tau_q$) and impact decay ($\tau_M$).}
    \label{fig:snr_linear}
\end{figure}

To further illustrate the robustness of this  effect, in Figure~\ref{fig:snr_linear} we vary the timescales $\tau_M$ and $\tau_q$ both in tandem. Higher $\tau_q$ allows for larger deviations from TWAP, which makes the noise reduction less efficient, but the improvement remains significant for realistic values of the parameters.

\section{Enhanced Estimation of Impact Cost}

\label{sec:impact}

In the preceding section we have developed an improved estimator of linear cost. Our next objective is to also obtain a good estimator for the magnitude $\lambda$ of the price impact.

\subsection{Removing the Linear Cost}

To this end, we first consider a straightforward metric, the total price impact of the order:
\begin{equation*}
    I = M_T - M_0.
\end{equation*}
In the Obizhaeva-Wang model this is given by
\begin{align*}
I = \int_0^T dM_t = \lambda \int_0^T q_te^{-(T-t)/\tau_M}dt + \sigma_M W^M_T.
\end{align*}
As a consequence, the expected total impact is
\begin{align}
   \mathbb{E}[I] &= \lambda \int_0^T\mathbb{E}\left[q_t\right]e^{-(T-t)/\tau_M}dt =\frac{\lambda Q\tau_M}{T}\left(1-e^{-T/\tau_M}\right).
   \label{eq:EIT}
\end{align}
Because impact is measured on mid prices rather than execution prices, the linear cost term disappears. The representation~\eqref{eq:EIT} therefore seems convenient to estimate $\lambda$ from data. 

However, the variance of this quantity is empirically still rather large, because the market impact of trading realistically sized orders represents only a few percent of daily volatility~\citep{toth2011anomalous}. In contrast, the final mid price is affected by the total market noise accumulated over the order~\citep{appendix}:
\begin{align*}
 \mathrm{Var}[I] &= \sigma_M^2T + \left(\frac{\lambda Q\tau_M}{T}\right)^2\left(\frac{2\sigma_q^2\tau_q^2}{\tau_M+\tau_q}+1\right) + O(\tau^3).
\end{align*}

\subsection{Optimal Weighting of Mid-Price Changes for TWAP}

\begin{figure}[tb]
    \centering
    \includegraphics[width=0.8\linewidth]{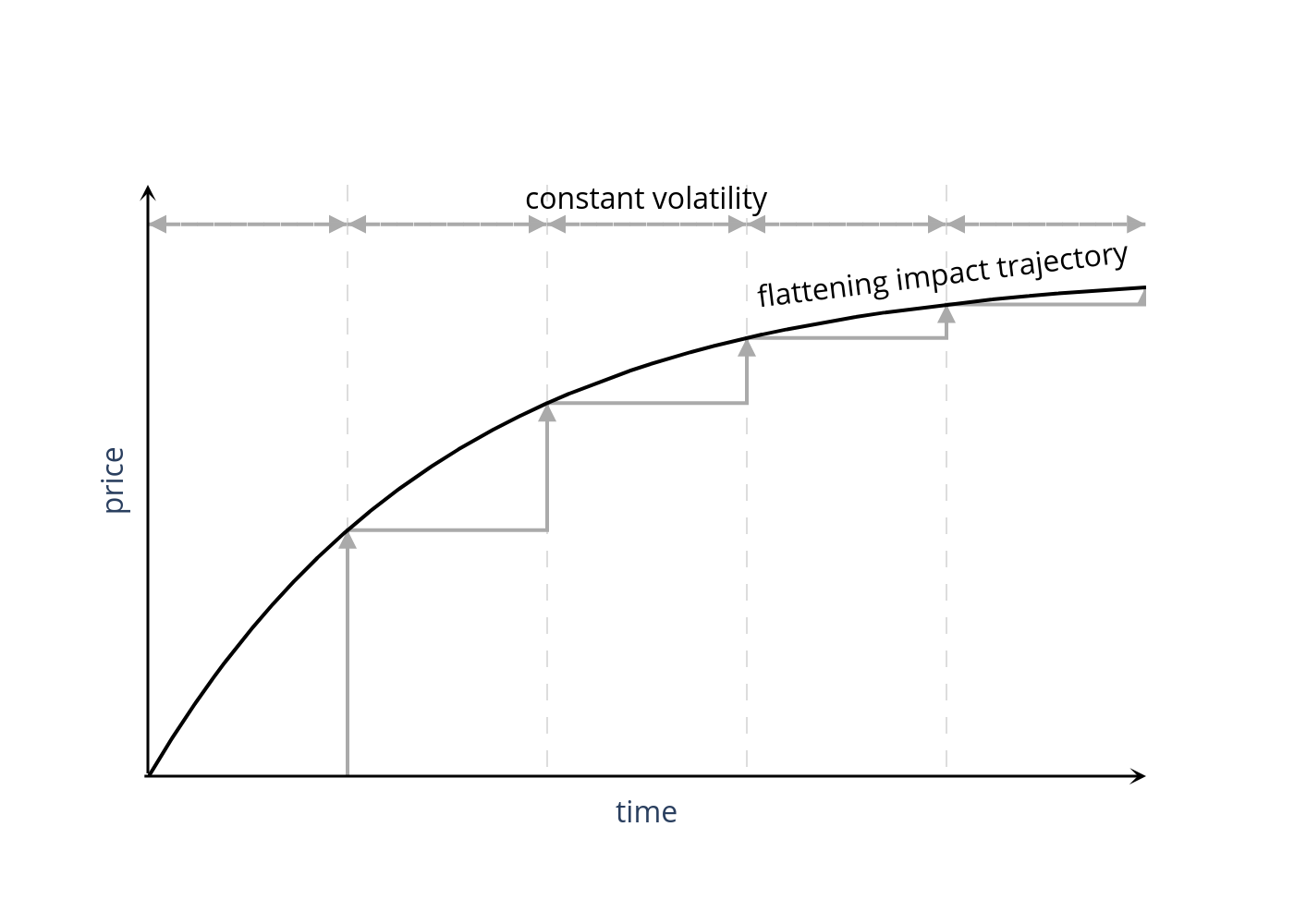}
    \caption{As a TWAP order progresses, the trajectory of cumulative impact flattens out, and the expected impact content of further returns decreases. Assuming a constant amount of market volatility, in an optimal estimator of price impact, later returns should gradually have lower weight.}
    \label{fig:impact_slices}
\end{figure}

To further improve the estimator, we recast it in the form
\begin{equation*}
I = \int_0^T \pi^I_t dM_t \quad \mbox{for the trivial weights} \quad \pi^I_t = 1.
\end{equation*}
Viewed in this way, estimating $\lambda$ from the total impact means that we look at an equal-weighted sum of price changes. This is sub-optimal because, as Eq.~\eqref{eq:EIT} shows, impact decay implies that the impact of a TWAP-like order exponentially converges to a constant level. Therefore, most of the price moves due to impact are concentrated into the early part of the order. Later on the impact trajectory flattens out, so by including these returns as well with the same weights we keep accumulating the same amount of market noise for ever less signal. This effect is illustrated in Figure~\ref{fig:impact_slices}.

We therefore look for a better set of weights $\pi_t$ that optimize the trade-off between signal and noise. To this end, define
\begin{equation*}
  I_\pi = \int_0^T \pi_t dM_t.
\end{equation*}
For the moment we focus on TWAP trading. Without fluctuations of the trading rate, we  can then choose the weights $\pi_t$ to be deterministic as well without loss of generality.

For a TWAP schedule the only stochastic term in the price process comes from the changes in the unaffected price driven by $W^M_t$. As a consequence,
\begin{eqnarray}
  \mathrm{Var}\left[I_\pi\right] = \mathrm{Var}\left[\int_0^T \sigma_M \pi_t dW^M_t\right] = \sigma_M^2 \int_0^T \pi_t^2 dt.
  \label{eq:varpi2}
\end{eqnarray}
In particular, for equal weights, we have $\mathrm{Var}[I] = \sigma_M^2T$. If we wish to choose the weights $\pi_t$ such that the variance of $I_\pi$ matches this value, Eq.~\eqref{eq:varpi2} turns into the constraint
\begin{eqnarray}
  \int_0^T \pi_t^2 dt = T.
  \label{eq:Q2T}
\end{eqnarray}
Among all the weights that produce this same level of variance, we can now try to maximize the expectation of the weighted price changes $I_\pi$. More specifically, maximizing the corresponding signal-to-noise ratio leads to the Lagrangian
\begin{eqnarray*}
    \mathcal{L}_\pi = \mathbb{E}\left[\int_0^T \pi_t dM_t \right] - \frac{\mu}{2}\left[\int_0^T \pi_t^2 dt - T\right]= \int_0^T \pi_t \frac{\lambda Q}{T}e^{-t/\tau_M} dt- \frac{\mu}{2}\left[\int_0^T \pi_t^2 dt - T\right].
\end{eqnarray*}
This resulting optimality condition is
\begin{eqnarray}
    \frac{\partial \mathcal{L}_\pi}{\partial \pi_t} = \frac{\lambda Q}{T}e^{-t/\tau_M}-\mu\pi_t dt = 0.
\end{eqnarray}
To pin down the Lagrange multiplier $\mu$ we use the constraint~\eqref{eq:Q2T}, obtaining the final solution 
\begin{align}
    \pi_t = \sqrt{2}\left(\frac{T}{\tau_M}\right)^{1/2}\left(1-e^{-2T/\tau_M}\right)^{-1/2}e^{-t/\tau_M}.
    \label{eq:pitTWAP}
\end{align}
These weights decay over time, which is in contrast with the equal weight approach of calculating total impact. The higher the contribution of impact at time $t$ to a local mid change, the higher we should be weighting it in our cost metric. This will help maximize the relative contribution of impact with respect to the market noise, whose volatility $\sigma_M$ was assumed to be constant across time. 

From the above equations we can express the expected signal with the optimized weights as
\begin{align}
    \mathbb{E}[I_\pi] & = \frac{\lambda Q}{\sqrt 2}\left(\frac{\tau_M}{T}\right)^{1/2}\left(1-e^{-2T/\tau_M}\right)^{1/2}.
    \label{eq:ecpitwap}
\end{align}
Comparing this to its counterpart~\eqref{eq:EIT} for equal return weights, we can deduce that for TWAP schedules, $\mathbb{E}[I]$ is of order $O(\tau_M/T)$, whereas $\mathbb{E}[I_\pi]$ is of order $O(
\sqrt{\tau_M/T})$. The latter can be much greater when impact decay is fast. On the other hand, the variance of the two quantities is equal. As a consequence, the signal-to-noise ratio of the estimator can be improved significantly by using front loaded rather than equal weights. 

\subsection{Optimal weighting of mid-price changes in the general case}

We now propose a generalization of the approach from the previous section to non-TWAP schedules. To this end recall that for deterministic TWAP trading rates,  the optimal weights satisfy
\begin{equation}\label{eq:weights}
\pi_t \propto \frac{d\mathbb{E}[M_t]}{dt}.
\end{equation}
For general stochastic trading rates, the expected mid price in the Obizhaeva-Wang model can still be computed \emph{conditionally} on the trading trajectory:
\begin{equation}
    \frac{\E[dM_t|q_{t'\in[0, T]}]}{dt} = \lambda q_t -\frac{\lambda}{\tau_M}\int_0^t e^{-(t-s)/\tau_M}q_s ds.
\end{equation}
In analogy to~\eqref{eq:weights}, we can then use the weights
\begin{equation}
    \pi_t = \nu \left(q_t - \frac{1}{\tau_M}\int_0^t ds e^{-(t-s)/\tau_M}q_s\right),
    \label{eq:pit}    
\end{equation}
where $\nu$ is a constant normalization factor. An opportune choice for the latter is 
\begin{equation*}
    \nu = \frac{\lambda Q T}{\mathbb{E}[C_\pi|q]},
    \label{eq:nusimplify}
\end{equation*}
because then $\mathrm{Var}[I] \approx \mathrm{Var}[I_\pi]$~\citep{appendix}. With this choice, we obtain
\begin{equation*}
    \mathbb{E}[I_\pi|q_{t'\in[0, T]}] = \lambda Q \left[\frac{\tau_M}{2T} + \sigma_q^2\tau_q \left(\frac{1}{2}+\frac{\tau_\mathrm{eff}^2}{T\tau_M}-\frac{\tau_\mathrm{eff}(\tau_M+3\tau_q)}{4T(\tau_M+\tau_q)}\right) \right]^{1/2} + O(e^{-T/\tau}).
\end{equation*}
As a sanity check, for TWAP schedules ($\sigma_q^2 = 0$), this formula simplifies to 
\begin{equation*}
    \mathbb{E}[I_\pi|\mathrm{TWAP}] = \lambda Q \left(\frac{\tau_M}{2T}\right)^{1/2},
\end{equation*}
which perfectly matches the leading-order term of the direct calculation~\eqref{eq:ecpitwap}.

\begin{figure}[tb]
    \centering
    \includegraphics[width=0.8\linewidth]{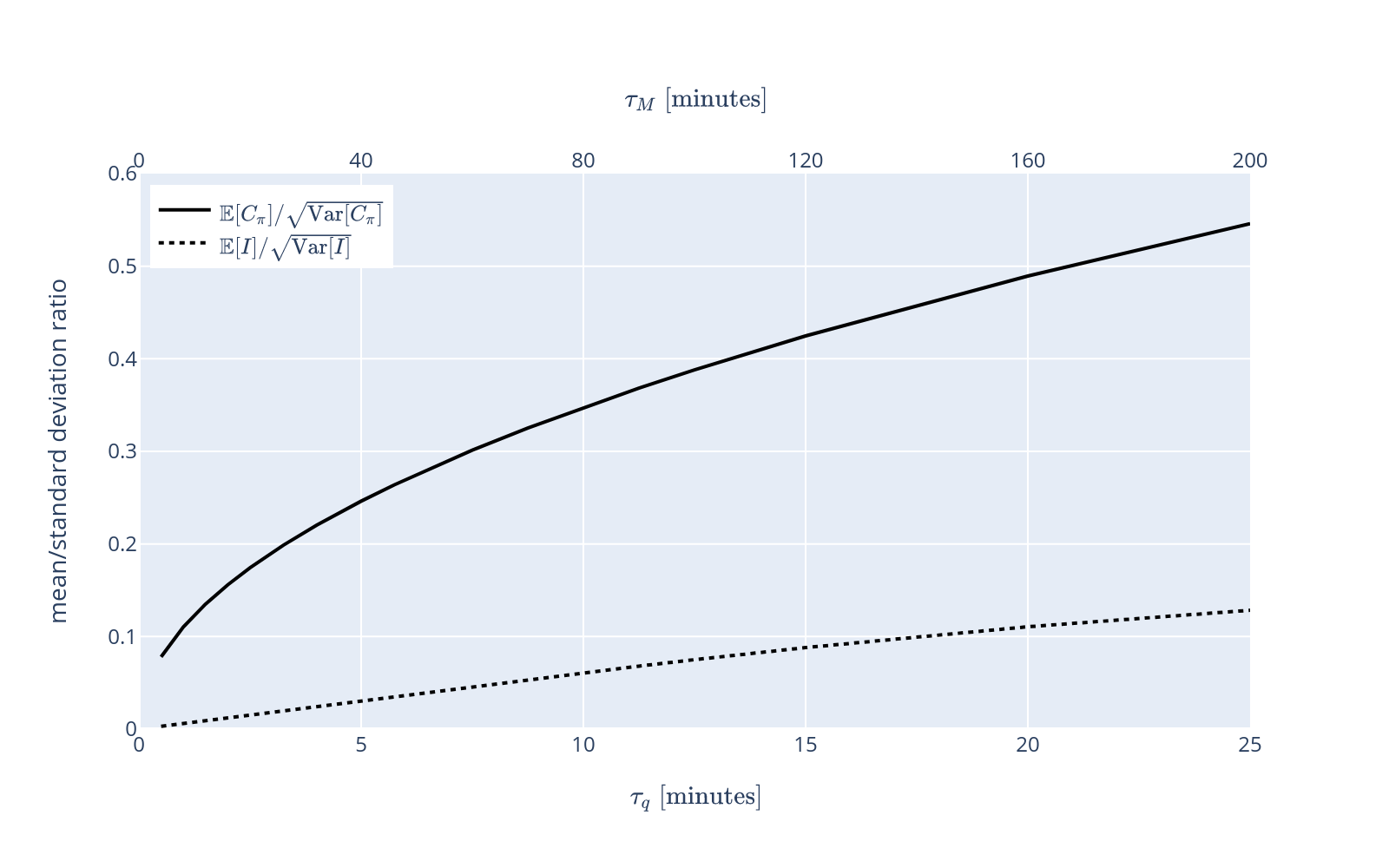}
    \caption{The ratio of the mean and standard deviation of the two impact cost metrics. The dotted line represents the naive metric $I$, impact to arrival price. The solid line represents $I_\pi$, the optimally weighted impact. The ratio of the two curves is close to a constant. On the horizontal axis we vary in tandem the time scales of trading rate fluctuations ($\tau_q$) and impact decay ($\tau_M$).}
    \label{fig:snr_impact}
\end{figure}

Table~\ref{tab:table_results} shows that, similarly as for linear costs, our approach allows to achieve a $7$ times improvement in signal-to-noise ratio relative to the na\"ive estimator. When measuring impact from $I_\pi$ instead of $I$ for a sample with $N=1,000$ orders from the unconditional simulation, the t-statistic of impact goes from $\sqrt{N} \mathbb{E}[I]/\sqrt{\mathrm{Var}[I]} \approx 0.95$ to $\sqrt{N}\mathbb{E}[I_\pi]/\sqrt{\mathrm{Var}[I_\pi]} \approx 7.32$. The table also shows, that the conditional and unconditional model yield very similar numerical results, both in line with the analytical approximations.

In Figure~\ref{fig:snr_impact}, we see that while for slower decay (greater $\tau_M$) the overall level of impact  increases. This affects both approaches equally, and the improvement of the signal-to-noise ratio remains comparable.

\section{Conclusion}
\label{sec:discussion}

This paper shows how to use intraday price and trading data to improve trading cost estimates. The results reported here for the model of \cite{obizhaeva2013optimal} can be generalized to more general models with transient price impact, such as the propagator models of~\cite{bouchaud2006random}. We expect our conclusions to hold in practice in a wide range of such scenarios.

Our objective is not to advocate for any specific model, but rather to propose a general principle for reducing noise in trading cost measurements. In fact, the two metrics that we propose, can be implemented even without relying on any model at all. We estimate linear cost via slippage to TWAP, which itself is defined using only observable, model-free quantities. As for impact cost, our enhanced weighting scheme does rely on knowing the impact profile during order execution. However, that has an empirically well documented and surprisingly universal shape~\citep{durin2023two}. This could be used as is, once more without imposing an explicit impact model. Our formulas thereby only leverage information readily available to most market participants, and are straightforward to implement or to adapt to the specific requirements of one's strategy.

Further research should explore various avenues to generalize this work. While the estimator of linear cost is not very sensitive to the specifics of price impact and its model, it can be extended beyond TWAP by adjusting the reference price to incorporate the average time profile of an arbitrary trading algorithm. Additionally, future studies could evaluate the performance of various impact models beyond Obizheava-Wang on real data for the weighted price impact estimator. This could lead to additional improvements in the signal-to-noise ratio in practice.

\bibliographystyle{abbrvnat}
\bibliography{m2_ab}

\end{document}